\documentclass[aps,prl,twocolumn,groupedaddress]{revtex4}

\usepackage {graphics}

\begin{document}


\title{A correction method for the pair density 
to get close to the ground state one}


\author{Masahiko Higuchi}
\affiliation{Department of Physics, Faculty of Science, Shinshu University, 
Matsumoto 390-8621, Japan}

\author{Katsuhiko Higuchi}
\affiliation{Graduate School of Advanced Sciences of Matter, 
Hiroshima University, Higashi-Hiroshima 739-8527, Japan}


\date{\today}

\begin{abstract}
We present a correction method for the pair density (PD) to get close to the 
ground state one. The PD is corrected to be a variationally-best PD within 
the search region that is extended by adding the uniformly-scaled PDs to its 
elements. The corrected PD is kept $N$-representable and satisfies the virial 
relation rigorously. The validity of the present method is confirmed by 
numerical calculations of neon atom. It is shown that the root-mean-square 
error of the electron-electron interaction and external potential energies, 
which is a good benchmark for the error of the PD, is reduced by 
69.7{\%} without additional heavy calculations.
\end{abstract}


\maketitle

%
The electron correlation is one of main topics in the fields of the 
atomic, molecular and condensed matter physics. 
The physical quantity that directly expresses the electron 
correlation is the diagonal element of the second-order reduced density 
matrix, which is so-called the pair density (PD). The PD gives not only the 
electron density but also expectation values of arbitrary two-particle 
operators\cite{1,2,3}. Therefore, 
the PD functional theory (PDFT)\cite{6,7,8,9,10,11,12,13,14,16,17,18,
19,20,21,23,26,27,28,30,32,33,35,36,37,38,39,40,41,42} 
is one of the promising first-principle methods for describing the electron 
correlation beyond the conventional 
density functional theory\cite{43, 44}.
In this letter, we present a correction method for the PD 
to get close to the ground state one, 
in which the search region of PDs is substantially extended. 
This method can become a powerful tool to overcome the well-known problems 
of the PDFT. 

There exist two kinds of well-known problems in developing the PDFT. 
One is to have to develop the approximate form of the kinetic energy functional
since the kinetic energy cannot be expressed by the PD 
alone\cite{1,2,3,39,40,41,42}. 
Some approximation is needed for the kinetic energy 
functional of the PDFT\cite{39,40,41,42}. 
The other problem is related to the $N$-representability problem 
of the PD\cite{1,2,3,26,27,28,30,32,33,35,36,37,38}. 
Although we have to search the ground state PD within the set of 
the $N$-representable PDs (2nd Hohenberg-Kohn (HK) theorem of the PDFT), 
the necessary and sufficient conditions for the 
$N$-representability of the PD have not yet been known 
in a practical form\cite{1,2,3,26,27,28,30,32,33,35,36,37,38}.

In this letter, concerning the first problem we propose an approximate 
functional of the kinetic energy along the strategy shown in the previous 
paper\cite{23}. 
Namely, an approximate functional is developed by imposing two kinds of 
sum rules for the kinetic energy functional as the restrictive conditions. 
Although the previous approximate functionals have been developed by 
imposing excessively the restrictive conditions on them\cite{23}, 
the present one is devised with removing such excessive restrictive 
conditions\cite{23,45}. We have

\begin{equation}
\label{eq1}
T\!\left[ {\gamma^{(2)}\!} \right] 
\!=\!\! \int \!\!\!\! 
\int \!\!{\left\{ { K 
{\frac{2}{r^{2}}} \!+\! {K}'\frac{\cos \theta 
}{r^{2}}\ln \!\left( {\frac{r}{{r}'}} \right)} \!\right\}}
\gamma^{(2)}({\rm {\bf r{r}'}}\!;\!{\rm {\bf r{r}'}})
\rm {d} \rm {\bf r} \rm {d} \rm{{\bf r}'},
\end{equation}

\noindent
where $\gamma^{(2)} ({\rm {\bf r{r}'}};{\rm {\bf r{r}'}})$ denotes the 
PD, and where $K$ and ${K}'$ are arbitrary constants that are determined 
later. Equation (\ref{eq1}) satisfies the scaling property of 
the kinetic energy functional and is consistent with the HK theorem of 
the PDFT\cite{23,45}. 
The former fact will be significant later 
in discussing a correction method.

Concerning the second problem, we have recently attempted to extend the 
variational search region of PDs\cite{20,23} beyond that of the effective 
initial theory\cite{19,21}. 
However, it has been eagerly anticipated that such 
a search region is further extended with keeping the $N$-representability of 
the PD. In this letter, we shall extend the search region by adding another 
set of PDs to the elements of the search region. This is precisely our 
correction method, and is a main subject of this letter. Let us show this 
method in the following three steps (i), (ii) and (iii).

(i) First of all, we shall show that the kinetic energy functional which is 
consistent with the scaling property satisfies the virial relation exactly. 
Here suppose that the search region of PDs fully covers a set of the 
$N$-representable PDs. Applying the scalings of the electron coordinates, 
$\gamma^{(2)} ({\rm {\bf r{r}'}};{\rm {\bf r{r}'}})$ is transformed into

\begin{equation}
\label{eq2}
\gamma_{\lambda }^{(2)} ({\rm {\bf r{r}'}};{\rm {\bf r{r}'}}) 
= \lambda^{6}\gamma^{(2)} 
(\lambda {\rm {\bf r}}\lambda {\rm {\bf {r}'}};
\lambda {\rm {\bf r}}\lambda {\rm {\bf {r}'}}),
\end{equation}

\noindent
where $\lambda $ is the scaling parameter. 
$\gamma_{\lambda}^{(2)}({\rm {\bf r{r}'}};{\rm {\bf r{r}'}})$ 
is the PD scaled from 
$\gamma^{(2)} ({\rm {\bf r{r}'}};{\rm {\bf r{r}'}})$, 
which is called the scaled PD. The kinetic energy functional 
that is consistent with the scaling property satisfies 
the following relation\cite{10,19,23}:

\begin{equation}
\label{eq3}
T\left[ {\gamma_{\lambda }^{(2)} } \right] 
= \lambda^{2}T\left[ {\gamma^{(2)} } \right].
\end{equation}

\noindent
Similarly, the electron-electron interaction and external potential 
energies to the scaled PD are, respectively, given by

\begin{equation}
\label{eq4}
W\left[ {\gamma_{\lambda }^{(2)} } \right] 
= \lambda W\left[ {\gamma^{(2)} } \right],
\end{equation}

\noindent
and

\begin{equation}
\label{eq5}
V\left[ {\gamma_{\lambda }^{(2)} } \right] = \frac{2}{N - 1}
\int\!\!\!\int 
{v_{ext} \left( {\frac{{\rm {\bf r}}}{\lambda }} \right)}
\gamma^{(2)} ({\rm {\bf r{r}'}};{\rm {\bf r{r}'}})
\rm {d} \rm {\bf r} \rm {d} \rm{{\bf r}'},
\end{equation}

\noindent
where $v_{ext} ({\rm {\bf r}})$ denotes the external potential and where $N$ 
is the number of electrons. 
Due to the 2nd HK 
theorem of the PDFT, the total energy functional $E\left[ 
{\gamma^{(2)} } \right]\left( { = T\left[ {\gamma^{(2)} } \right] + 
W\left[ {\gamma^{(2)} } \right] + V\left[ {\gamma^{(2)} } \right]} 
\right)$ is minimum at $\gamma_{0}^{(2)} $ that corresponds to the 
variationally-best PD among all of the $N$-representable PDs. Using the 
scaled PD, the theorem can be written as $\left. {{dE\left[ {\gamma 
_{0,\,\lambda }^{(2)} } \right]} \mathord{\left/ {\vphantom {{dE\left[ 
{\gamma _{0,\,\lambda }^{(2)} } \right]} {d\lambda }}} \right. 
\kern-\nulldelimiterspace} {d\lambda }} \right|_{\lambda = 1} = 0$. Here 
$\gamma_{0,\,\lambda }^{(2)} $ is the PD scaled from $\gamma_{0}^{(2)} $, 
and is defined similarly to Eq. (\ref{eq2}). 
Substituting Eqs. (\ref{eq3}), (\ref{eq4}) and (\ref{eq5}) 
into this theorem, we have

\begin{equation}
\label{eq6}
2T\!\left[ {\gamma_{0}^{(2)} } \right] 
\!+\! 
W\!\left[ {\gamma_{0}^{(2)} } \right] 
\!\!= \!\!
\int \!\!\!\! \int\!\!
\frac{\left\{ {{\rm {\bf r}} \!\cdot\! \nabla v_{ext} ({\rm 
{\bf r}})} \right\}}{N - 1}
{\gamma_{0}^{(2)} ({\rm {\bf r{r}'}}\!;\!
{\rm {\bf r{r}'}})}
\rm {d} \rm {\bf r} \rm {d} \rm{{\bf r}'}.
\end{equation}

\noindent
Equation (\ref{eq6}) is the virial relation of the PDFT. 
It should be noted that 
even though the kinetic energy functional that 
satisfies Eq. (\ref{eq3}) is not an exact but an approximate form, 
Eq. (\ref{eq6}) exactly holds if the search region 
fully covers the set of the $N$-representable PDs. 

(ii) Next, we shall consider the case where the search region not 
fully but partially covers a set of the $N$-representable PDs. Suppose that 
such a partial search region is denoted as $C_{1} $ (See, Fig. 1), 
and that the total energy functional is minimum at $\tilde {\gamma 
}_{0}^{(2)} $ that corresponds to the variationally-best PD within the set 
$C_{1} $. Due to the incomplete cover of $N$-representable PDs by the set 
$C_{1} $, it could be that not all of the scaled PDs $\tilde {\gamma 
}_{0,\,\lambda }^{(2)} $ are included in the set $C_{1} $. We denote the set 
of $\tilde {\gamma }_{0,\,\lambda }^{(2)} $ as $C_{2}$ 
(Fig. 1). In this case, the total energy functional 
$E\left[ {\tilde {\gamma }_{0,\,\lambda }^{(2)} } \right]$ 
does not always take the 
minimum at $\lambda = 1$. Namely, we have 
$\left. {{dE\left[ {\tilde {\gamma }_{0,\,\lambda }^{(2)} } \right]} 
\mathord{\left/ {\vphantom {{dE\left[ {\tilde {\gamma }_{0,\,\lambda }^{(2)} 
} \right]} {d\lambda }}} \right. \kern-\nulldelimiterspace} {d\lambda }} 
\right|_{\lambda = 1} \ne 0$.
%
%
%
%
Like the derivation of Eq. (\ref{eq6}), this inequality 
leads to the relation such that the 
left-hand side of Eq. (\ref{eq6}) is \textit{not} equal to 
the right-hand side. 
Thus, if the search region 
does not fully cover the set of the $N$-representable PDs, the 
virial relation does \textit{not} hold 
even though the kinetic energy functional 
satisfies Eq. (\ref{eq3}). This fact can be used as a criterion of whether the 
search region fully covers the set of the $N$-representable PDs or not. Here 
let us define the virial ratio which indicates to what extent the virial 
relation holds:

\begin{equation}
\label{eq8}
R_{v} \!=\! 
\frac{W\!\left[ {\tilde {\gamma}_{0}^{(2)} } \right] \!-\! 
\displaystyle \int\!\!\!\!\int 
\frac{\left\{ {{\rm {\bf r}} \!\cdot\! \nabla v_{ext} 
({\rm {\bf r}})} \right\}}{N-1}
{\tilde {\gamma }_{0}^{(2)} 
({\rm {\bf r{r}'}}\!;\!{\rm {\bf r{r}'}})} 
\rm {d} \rm {\bf r} \rm {d} \rm{{\bf r}'}}
{T\left[ {\tilde 
{\gamma }_{0}^{(2)} } \right]}.
\end{equation}

\noindent
Using this ratio, the above-mentioned criterion can be restated as follows; 
\textit{the deviation of }$R_{v}$
\textit{ from the value -2.0 means the insufficiency of 
the search region of PDs if the kinetic energy functional 
satisfies Eq. (\ref{eq3})}\cite{47}. 

(iii) When $R_{v} $ deviates from -2.0, a correction method that is 
related to the extension of the search region is desired. The key point of 
the present correction method is the extension of the search region 
by adding the scaled PDs to the search region. Figure 1 truly shows the 
relation between the original search region $C_{1} $ and newly-added search 
region $C_{2} $. 
Both sets $C_{1} $ and $C_{2}$ are subsets of the set of 
the $N$-representable PDs. 
The above-mentioned inequality $\left. {{dE\left[ {\tilde {\gamma 
}_{0,\,\lambda }^{(2)} } \right]} \mathord{\left/ {\vphantom {{dE\left[ 
{\tilde {\gamma }_{0,\,\lambda }^{(2)} } \right]} {d\lambda }}} \right. 
\kern-\nulldelimiterspace} {d\lambda }} \right|_{\lambda = 1} \ne 0$ 
means that the set $C_{2}$ 
could possibly include the PD that takes a total energy lower than 
what the best PD within the set $C_{1} $, i.e., 
$\tilde {\gamma }_{0}^{(2)} $, takes. 
Such a PD exists in $C_{2} \cap \bar {C}_{1} $ and is denoted as 
$\tilde {\gamma }_{0,\,\Lambda }^{(2)} $, 
where $\bar {C}_{1} $ is the complementary set of $C_{1} $ (Fig. 1). 
Then, the inequality 
$\left. {{dE\left[ {\tilde {\gamma }_{0,\,\lambda 
}^{(2)} } \right]} \mathord{\left/ {\vphantom {{dE\left[ {\tilde {\gamma 
}_{0,\,\lambda }^{(2)} } \right]} {d\lambda }}} \right. 
\kern-\nulldelimiterspace} {d\lambda }} \right|_{\lambda = 1} \ne 0$ 
is rewritten as 
$\left. {{dE\left[ {\tilde {\gamma }_{0,\,\lambda }^{(2)} } 
\right]} \mathord{\left/ {\vphantom {{dE\left[ {\tilde {\gamma 
}_{0,\,\lambda }^{(2)} } \right]} {d\lambda }}} \right. 
\kern-\nulldelimiterspace} {d\lambda }} \right|_{\lambda = \Lambda } = 0$.
%
%
%
%
Substituting Eqs. (\ref{eq3}), (\ref{eq4}) and (\ref{eq5}) 
into this equation, we obtain

\begin{equation}
\label{eq10}
2\Lambda T\!\left[ {\tilde {\gamma }_{0}^{(2)} } \!\right] 
\!+ W\!\left[ {\tilde {\gamma }_{0}^{(2)} } \!\right] 
\!\!=\!\!\!
\int \!\!\!\!\! \int \!\!
\frac{\left\{ 
{{\rm {\bf r}} \!\cdot\!\! 
\nabla\!_{\frac{\rm {\bf r}}{\Lambda}} 
v_{ext}(\frac{\rm {\bf r}}{\Lambda})} \right\}}{\Lambda ^{2}(N - 1)}
\tilde {\gamma}_{0}^{(2)} 
({\rm {\bf r{r}'}}\!;\!{\rm {\bf r{r}'}})
\rm {d} \rm {\bf r} \rm {d} \rm{{\bf r}'}.
\end{equation}

\noindent
If we consider the isolated atomic system, 
Eq. (\ref{eq10}) is easily rewritten as 

\begin{equation}
\label{eq11}
2\Lambda{T\left[ {\tilde {\gamma }_{0}^{(2)} } \right]}
+ {W\left[ {\tilde {\gamma }_{0}^{(2)} } \right] =
- V\left[ {\tilde {\gamma }_{0}^{(2)} } \right]}.
\end{equation}

\noindent
From Eq. (\ref{eq10}) or (\ref{eq11}), we can determine 
the value of $\Lambda $. The scaled PD 
$\tilde {\gamma}_{0,\,\Lambda }^{(2)} $ with $\Lambda $ 
thus determined is the variationally-best one 
within the union of two sets, i.e., 
$C_{1} \cup C_{2} $. 
That is to say, a correction of the ground state PD from 
$\tilde {\gamma }_{0}^{(2)} $ to $\tilde {\gamma }_{0,\,\Lambda }^{(2)} $ is 
accomplished by extending the search region from 
$C_{1} $ to $C_{1} \cup C_{2} $. 
It is obvious that the corrected PD 
$\tilde {\gamma }_{0,\,\Lambda}^{(2)} $ 
remains $N$-representable. Furthermore, it is easily confirmed 
that the virial relation exactly holds for the corrected PD 
$\tilde {\gamma}_{0,\,\Lambda }^{(2)} $.

\begin{figure}[htbp]
\scalebox{0.5}{\includegraphics{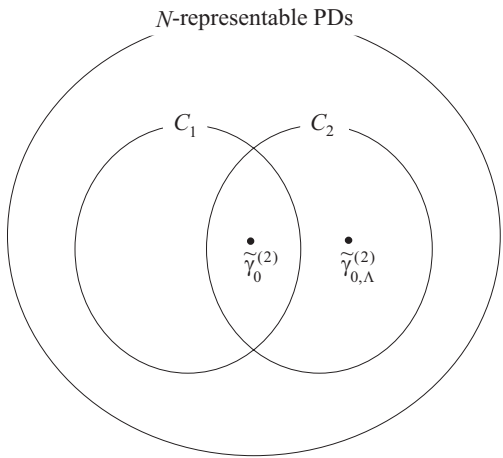}}
\caption{\label{fig1} The relations between various sets of PDs. The set $C_{1}$ covers a set of the $N$-representable PDs not fully but partially. The set $C_{2} $ consists of the uniformly-scaled PDs. The variationally-best PD within $C_{1} $, and that within $C_{1} \cup C_{2} $, are denoted by $\tilde {\gamma }_{0}^{(2)} $ and $\tilde {\gamma }_{0,\Lambda }^{(2)} $, respectively. }
\end{figure}

The procedure of this correction method is as follows. 
First we prepare the approximate form of the kinetic energy functional 
that satisfies Eq. (\ref{eq3}), 
and determine the search region of PDs. Using this kinetic energy 
functional, we calculate 
$\tilde {\gamma }_{0}^{(2)} $. 
Substituting $\tilde {\gamma }_{0}^{(2)} $ 
into Eq. (\ref{eq10}) or (\ref{eq11}), the scaling parameter 
$\Lambda$ is obtained. 
This $\Lambda $ immediately leads to the corrected PD 
$\tilde {\gamma}_{0,\,\Lambda }^{(2)}$ 
by using Eq. (\ref{eq2}). 
Of course, the kinetic energy, electron-electron interaction energy 
and external potential energy for 
$\tilde {\gamma }_{0,\,\Lambda }^{(2)} $ 
can be easily calculated from 
Eqs. (\ref{eq3}), (\ref{eq4}) and (\ref{eq5}), respectively. 
One of striking points of this correction method is 
that the heavy calculation task is \textit{not} needed for the 
corrections of the PD and related quantities.
%

In order to check the validity of the present method, we perform 
numerical calculations for neutral neon atom. We utilize the computational 
PD functional scheme\cite{23} so as to obtain $\tilde {\gamma }_{0}^{(2)} $. 
According to this scheme\cite{23}, 
we adopt as the search region the set of PDs 
that are calculated from the correlated wave functions. As the correlated 
wave functions, we take the linear combination of the ground state and 
doubly-excited Slater determinants (SDs) that consist of eigenfunctions 
of the effective initial scheme\cite{19,21}. 
Among all doubly-excited SDs, we 
choose the SDs that have non-negligible contributions to the resultant PD. A 
total of 2,861 SDs and 12,275 SDs, the constituent eigenfunctions of which 
have the principal quantum number up to 6 and 11, respectively, are chosen 
as basis functions in constructing PDs.

As the set $C_{1} $, we take the set of PDs that are constructed by 2,861 
SDs. First, by means of the computational PD functional scheme\cite{23}, 
we calculate $\tilde {\gamma }_{0}^{(2)} $ that is the 
variationally-best PD within the set $C_{1} $. Then, the corrected PD 
$\tilde {\gamma }_{0,\,\Lambda }^{(2)} $ 
is calculated by using Eqs. (\ref{eq2}) and (\ref{eq11}). 
We also prepare the search region that is extended with 
increasing the number of basis SDs (NSD) from 2,861 to 12,275. Hereafter, we 
denote this search region of PDs as ${C}'_{1} $. For comparison, the 
variationally-best PD within the set ${C}'_{1}$ is also calculated by means 
of the computational PD functional scheme\cite{23}. 

Using $\tilde {\gamma }_{0}^{(2)} $ and 
$\tilde {\gamma }_{0,\,\Lambda }^{(2)}$, 
we calculate the errors of the electron-electron interaction energy, 
external potential energy, kinetic energy and virial ratio that are 
denoted by $\Delta W$, $\Delta V$, $\Delta T$ and $\Delta R_{v} $, 
respectively. The root-mean-square error (RMSE) of the electron-electron 
interaction and external potential energies is also calculated by $\sqrt 
{\left\{ {(\Delta W)^{2} + (\Delta V)^{2}} \right\} / 2} $. The RMSE is 
considered as a good benchmark for to what extent the resultant PD is close 
to the correct ground state PD. This is because accuracies of both the 
electron-electron interaction and external potential energies are dependent 
only on that of the PD. 

The above-mentioned calculations are performed with changing the values of 
$K$ and ${K}'$ that appear in Eq. (\ref{eq1}). These values are determined by 
requiring the approximate functional to have two desirable features. One is 
that the approximate functional reproduces the Hartree-Fock kinetic energy 
if the PD coincides with the Hartree-Fock PD. The other is that the RMSE of 
the electron-electron interaction and external potential energies is 
minimized with respect to parameters. This determination process is 
implemented in individual cases before and after the correction. 
%

Calculation results are shown in Fig. 2. From this figure, we find the 
following points. 

(I) It is found in Fig. 2 that the present method reduces $\Delta W$ and 
$\Delta V$ by 89.7{\%} and 66.3{\%}, respectively 
(reductions from case $a$ to $c$ for 
``$\Delta W$'' and ``$\Delta V$'', respectively, in Fig.2). 
On the other hand, only 1.3{\%} and 1.1{\%} reductions are made 
via increasing the NSD from 2,861 to 12,275 in 
$\Delta W$ and $\Delta V$, respectively 
(reduction from case $a$ to $b$ for ``$\Delta W$'' and ``$\Delta V$'', 
respectively). 
Also, as shown in Fig. 2, the RMSE is remarkably reduced by 69.7{\%} 
(reduction from case $a$ to $c$) , while it is improved only 1.2{\%} 
by increasing the NSD (reduction from case $a$ to $b$). 
Thus, the reduction rate of the present method is much larger than that 
of the extension method with increasing the NSD. This leads to that the search 
region of PDs is more effectively extended by adding the uniformly-scaled 
PDs than by increasing the NSD. Furthermore, it should be noticed that 
the present method is feasible without additional heavy calculations.

\begin{figure}[htbp]
\scalebox{0.55}{\includegraphics{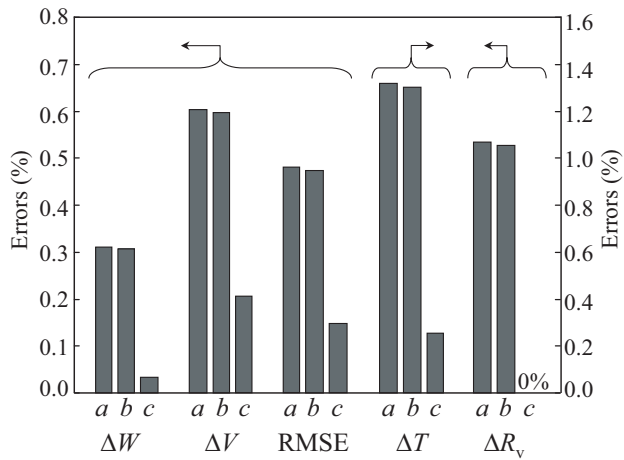}}
\caption{\label{fig2} 
The calculation results of the $\Delta W$, $\Delta V$, RMSE, 
$\Delta T$ and $\Delta R_{v} $. 
In this figure, the search region $C_{1} $, 
${C}'_{1} $ and $C_{1} \cup C_{2} $ are simply denoted 
by the symbols "$a$", "$b$" and "$c$", respectively. }
\end{figure}

(II) It is found in Fig. 2 that $\Delta T$ is also reduced much more 
effectively by the present method than by increasing the NSD (80.9{\%} 
and 1.2{\%} reductions for the former and latter, respectively). In general, the accuracy of the 
kinetic energy is dependent on both the appropriateness of the approximate 
form of the functional and that of the resultant PD. Judging from the 
results of the RMSE, $\Delta W$ and $\Delta V$ (Fig. 2), the resultant PD 
seems to be improved to be close to the ground state one. Therefore, we 
can deduce that the approximate form itself, which is given by Eq. (1), 
would be also sound.

(III) It is found in Fig. 2 that the virial ratios $R_{v} $ before the 
correction deviate from the correct value -2.0. Using the criterion on the 
search region, these deviations mean that 
not only the search region $C_{1} $ but also the search region ${C}'_{1} $ 
are insufficient since the approximate functional Eq. (\ref{eq1}) is consistent with 
Eq. (\ref{eq3}). That is to say, even though more than 12,000 SDs are used in 
constructing PDs, the search region is not extended effectively. On the 
other hand, $R_{v} $ after the correction is rigorously equal to -2.0, 
as it should be. These tendencies seem to be consistent with those mentioned 
in the above (I) and (II).

Thus, the present method improves not only $\Delta W$, $\Delta V$ 
and RMSE but also $\Delta T$ and $\Delta R_{v}$ quite substantially and 
effectively. We can therefore say that the PD is corrected to 
be close to the ground-state one appropriately via the present method. 

%

In conclusion, we summarize the features of the correction method proposed 
here. The most distinctive feature is that

\noindent
(1) the search region is extended by adding a set of the scaled 
PDs to elements of the search region. 

\noindent
The resultant PD that is corrected by this method possesses the following 
features: 

\noindent
(2) the corrected PD is $N$-representable,

\noindent
(3) the corrected PD satisfies the virial relation exactly.

\noindent
The validity of the present method is successfully confirmed by 
numerical calculations of neon atomic system. It is shown that

\noindent
(4) not only the RMSE but also the errors of the kinetic energy, 
electron-electron interaction energy and external potential energy are 
individually all reduced definitely, 
%
%

\noindent
(5) the correction is enough effective even in the small size calculations.

This correction method provides an ingenious way to extend the search region 
of PDs efficiently. Due to the computational easiness of the correction, 
this method will be extensively applicable to larger systems such as 
molecules, clusters and solids.

\end{document}